\documentclass[pdflatex,sn-mathphys]{sn-jnl}

\usepackage{amsmath,amssymb,amsfonts}
\usepackage{graphicx}
\usepackage{textcomp}
\usepackage{xcolor}
\usepackage{caption}
\usepackage{subcaption}
\usepackage{gensymb}
\usepackage[T1]{fontenc}

\jyear{2021}%

\begin{document}

\title[]{Increasing Physical Layer Security through Hyperchaos in VLC Systems}


\author[1]{\fnm{Ashish} \sur{Sharma}}\email{ashish2309.official@gmail.com}

\author*[1]{\fnm{Harshil} \sur{Bhatt}}\email{harshilbhatt2001@gmail.com}

\affil[1]{\orgdiv{Department of Electronics and Communication Engineering}, \orgname{Manipal Institute of Technology}, \orgname{Manipal Academy of Higher Education}, \orgaddress{\city{Manipal}, \state{Karnataka}, \country{India}}}



\abstract{
Visible Light Communication (VLC) systems have relatively higher security compared with traditional Radio Frequency (RF) channels due to line-of-sight (LOS) propagation. However, they still are susceptible to eavesdropping. 
The proposed solution of the papers have been built on existing work on hyperchaos-based security measure to increase physical layer security from eavesdroppers.
A fourth-order Henon map is used to scramble the constellation diagrams of the transmitted signals. The scramblers change the constellation symbol of the system using a key. That key on the receiver side de-scrambles the received data.
The presented modulation scheme takes advantage of a higher degree of the map to isolate the data transmission to a single dimension, allowing for better scrambling and synchronization.
A sliding mode controller is used at the receiver in a master-slave configuration for projective synchronization of the two Henon maps, which helps de-scramble the received data. 
The data is only isolated for the users aware of the key for synchronization, providing security against eavesdroppers. 
The proposed VLC system is compared against various existing approaches based on various metrics.
An improved Bit Error Rate and a lower information leakage are achieved for a variety of modulation schemes at an acceptable Signal-to-Noise Ratio.  
}

\keywords{visible light communication, sliding mode control, physical layer security, chaos synchronization, constellation scrambling}


\pacs[MSC Classification]{94A60, 68P25}
\maketitle

\section{Introduction}
\label{section:Introduction}
Optical wireless communication is considered a promising technology for indoor wireless communication due to the virtually unlimited and unlicensed spectrum of visible light. 
Due to this, Visible Light Communication (VLC) has had interest from researchers. 
In this form of wireless communication, an array of white light-emitting diodes (LEDs) is used as a transmitter, while a photodiode acts as a receiver.
VLC has several promising features which complement the traditional Radio Frequency (RF) wireless communication.
VLC systems are unaffected by the interference caused due to RF-based networks. They can also supplement radio-based networks where RF signals are significantly attenuated \cite{Blinowski2019SecuritySurvey}. 
VLC systems have a lower cost since the need for a carrier is eliminated as most indoor and outdoor environments are well illuminated.
VLC is considered a safer technology compared to radio-based technology as it does not generate RF multi-band interference \cite{Blinowski2019SecuritySurvey}. 
Hence, VLC based technology can be utilized in hazardous environments such as petroleum refineries and applications where WiFi communication causes interference, such as aircraft cabins.

Security requirements for VLC are similar to general security requirements for all wireless communication. 
Due to noisy environments, signal superposition may occur, leading to distortion and causing a drop in the authenticity and integrity of the signal at the receiver end.
Since a VLC transmitter broadcasts data, shielding of the information from eavesdroppers is a significant concern \cite{Lu2015High-SecurityShifting}. 
The coherence of light limits the throughput of VLC systems. The coherent length defines the minimum distance from which meaningful phase information can be retrieved. This coherence distance is \(20\mu\)m for LEDs, making phase modulation inappropriate for VLC \cite{Classen2016OpportunitiesLayer}. 

Existing research lies in masking the transmitted data with a chaos map. This is discussed in more detail in Section \ref{section:Literature Review}.
We propose a novel hyperchaotic VLC system to improve the system's security from an eavesdropper. Chaotic systems are highly sensitive to the initial value, never repeating and exhibit a non-periodic long term behaviour. This provides an advantage while transmitting data. 
We use a four-dimensional (4D) hyperchaotic Henon map to generate chaos which is used to scramble the constellation of the transmitted data. 
A sliding mode controller is used for chaos synchronization. 
The system is designed as a DC Optical Orthogonal Frequency-Division Multiplexing (DCO-OFDM) modulation using Intensity Modulated/Direct Detection (IM/DD).

The proposed solution utilizes a 4D Henon map to generate hyperchoas in the transmitter providing users 4 dimensions of data, which can be used independently to implement algorithms to improve physical layer security of any given system. The paper contributes in presenting results of a multi-dimensional hyperchaotic map in a VLC channel. It also presents a novel approach of scrambling constellations of  transmitted data and the effect of multiple layers of scrambling on BER and information leakage of a system. We achieve satisfactory BER and throughput, indicating increased physical layer security in VLC against eavesdroppers.

The rest of the paper is organized as follows: Section \ref{section:Literature Review} summarises relevant work followed by Section \ref{section:4D Hyperchaotic Henon Map} which describes the hyperchaotic map, the chaos sequence generation and synchronization. Section \ref{section:System Design} describes the proposed system followed by the description of the communication channel in Section \ref{section:Communication Channel}. The simulation results are discussed in Section \ref{section:Simulation}. Finally, the paper is concluded with Section \ref{section:Conclusion}. 

\section{Literature Review}
\label{section:Literature Review}

 In recent years, \cite{Elgala2011, Elgala2009, Kumar2010} have been the first papers to survey LED-based VLC systems with their applications \cite{Sevincer2013}. brings attention to LEDs ability to provide smart-lighting and data transmission simultaneously. Similarly, \cite{Wu2014} expands on the benefits and challenges of VLC in the comparison of RF networks. The authors in \cite{Karunatilaka2015} investigated various methods for enhancing the performance of VLC, like modulation techniques and dimming control techniques. The authors in \cite{Qiu2016} concentrated on different channel models for VLC, whereas \cite{Sindhubala2016} inspected noise sources and noise mitigation methods for VLC. In a recent comparative study between a user-centric and network-centric design for VLC channels, various positioning and localization techniques were reviewed for indoor and outdoor VLC applications \cite{Zhuang2018}. On the other hand, a survey reported all the optimization techniques aimed to improve the performance of VLC systems. The emphasis was given to new technologies like Non-Orthogonal Multiple-Access (NOMA), Simultaneous Wireless Information and Power Transfer (SWIPT), cooperative transmission and Space Division Multiple Access (SDMA) \cite{Obeed2019}.

While VLC systems provide better security than RF-based technology, security is still a significant concern due to the open nature of this technology. 

Leung-Yan-Cheong et al. started the notion of secrecy capacity \cite{Parker}. After which, Oggier et al. derived the secrecy capacity expressions for wiretap channels, assuming Both the legitimate user's and eavesdropper's Channel State Information (CSI) is known \cite{Oggier2011}.

VLC based technologies work do not penetrate walls like RF, improving security. Claasen et al. show eavesdropping is still possible, even from outside the room, using door gaps, windows, keyholes or hiding eavesdropping devices in the room \cite{Classen2015TheSN}. Mostafa et al. propose a solution establishing secure communication zones where eavesdropping is physically impossible \cite{Wang2018Physical-layerAnalysis}. However, this reduces the range of the VLC transmitter. Lapodoth et al. formulate a theoretical rationale for jamming VLC systems \cite{Lapidoth1998TheChannel}. For a system at peak intensities, a channel with limited capacity and sufficient transmitting power, the channel will saturate and obscure the data source. This is also achievable with multiple rogue low-power transmitters.

Chaos is defined as aperiodic, long-term behaviour in a deterministic system that exhibits sensitive dependence on initial conditions.
A chaotic system needs to be aperiodic, unlike an unstable system where infinity works as a fixed point the system is repelled towards. A dynamic system must be sensitive to initial conditions, topologically transitive and have dense periodic orbits to be classified as chaotic.   
Hyperchaos is defined by its characteristic of having one or more positive Lyapunov exponent than chaos, providing greater security in communication \cite{Strogatz2018}.

Although the infusion of chaos in Visible Light Communication is novel, it has become more prevalent. There have been different approaches for gaining better security, such as channel scrambling and modulated M-Quadrature Amplitude Modulation (M-QAM) Mapper \cite{XiaogeWu2016}. A substantial drawback is the high Bit Error Rate (BER) caused by adding chaos maps, making a VLC system unsuitable for regular use.

Lee et al. analyze different chaos maps and propose a chaos map that has anti-jamming characteristics \cite{Ryu2013}. Sadoudi et al. proposed a robust additive hyperchaotic masking algorithm and implemented it on FPGAs \cite{Sadoudi2013}. A significant drawback is that the system is only suitable for wired digital communication.
Fu et al. propose a hyperchaos based medical image protection scheme. An extensive analysis is also carried out highlighting the security of the approach \cite{Fu2015}. Hammami et al. present a third-order Henon map to encrypt one-dimensional audio signals \cite{Hammami2016}. 
Yang et al. implement a fractional-order hyperchaotic system and numerically simulate the synchronization. The authors make use of adaptive control and fractional order stability to implement an adaptive self-synchronization controller \cite{Yang2015}.

Gunawan et al. propose and implement a Colour-Shift Keying (CSK) OFDM, using RGB LEDs, providing three channels for communication \cite{Gunawan2020}. Using three channels instead of a single channel from a regular LED helps increase the system's data rate. Chen et al. improve this work by using a chaos-based channel scrambling scheme capable of changing the communication channel dynamically. This improves the security of the system \cite{Chen2016}. 
A notable drawback of this proposal is the increased BER, which makes it unsuitable for use at high data rates or regular use. Fang et al. propose a chaotic generator based on the Gram-Schmidt Normalisation to reduce the BER using CSK \cite{Fang2017}.  

Most research to increase VLC systems' security involves masking the transmitted data with a chaos map. Based on the analysis above, a novel system in which a hyperchaos map is used to scramble the constellation diagrams of the transmitted data is proposed. 
The transmitter and receiver have different initial states for the hyperchaos generator, and a sliding mode controller is used to synchronize them.
Our approach builds upon existing research summarised above, along with a novel approach of scrambling constellations to increase security.

\section{4D Hyperchaotic Henon Map} 
\label{section:4D Hyperchaotic Henon Map}
In the proposed system, a 4D Henon map is used to drive the hyperchaotic transmitter. The transmitter and receiver are set at different initial values, and a sliding mode controller is used to synchronize them. The hyperchaotic map scrambles the transmitted data, providing security from eavesdroppers. 

A Henon map is a discrete-time dynamic system that maps two points into themselves. 
We use a 4D Henon map as described by the \(4^{th}\) order system of difference equation \cite{Richter2002TheChaos}.

\begin{equation}
  \label{eq:henon_map}
\begin{gathered}
x_1(k+1)  =a-x_3^2(k)+-bx_4(k)
\\
x_2(k+1)  =x_1(k)
\\
x_3(k+1)   =x_2(k)
\\
x_4(k+1)   =x_3(k)
\end{gathered}
\end{equation}

Where, \(x_1(k)\), \(x_2(k)\), \(x_3(k)\) and \(x_4(k)\) are the state variables and \(a\) and \(b\) are the system parameters. 

Let \(X_c =[x_1,x_2,x_3,x_4]^T\).
Equation \ref{eq:henon_map} exhibits hyperchaotic behaviour when system parameters \(a=1.76\) and \(b=0.1\) \cite{AbdulRahman2019SecureSystems}. The trajectories for the 4D Henon map with the aforementioned system parameters are plotted in Figure \ref{fig:trajectory_henon_map}.

\begin{figure}
     \centering
     \begin{subfigure}[b]{0.45\textwidth}
         \centering
         \includegraphics[width=\textwidth]{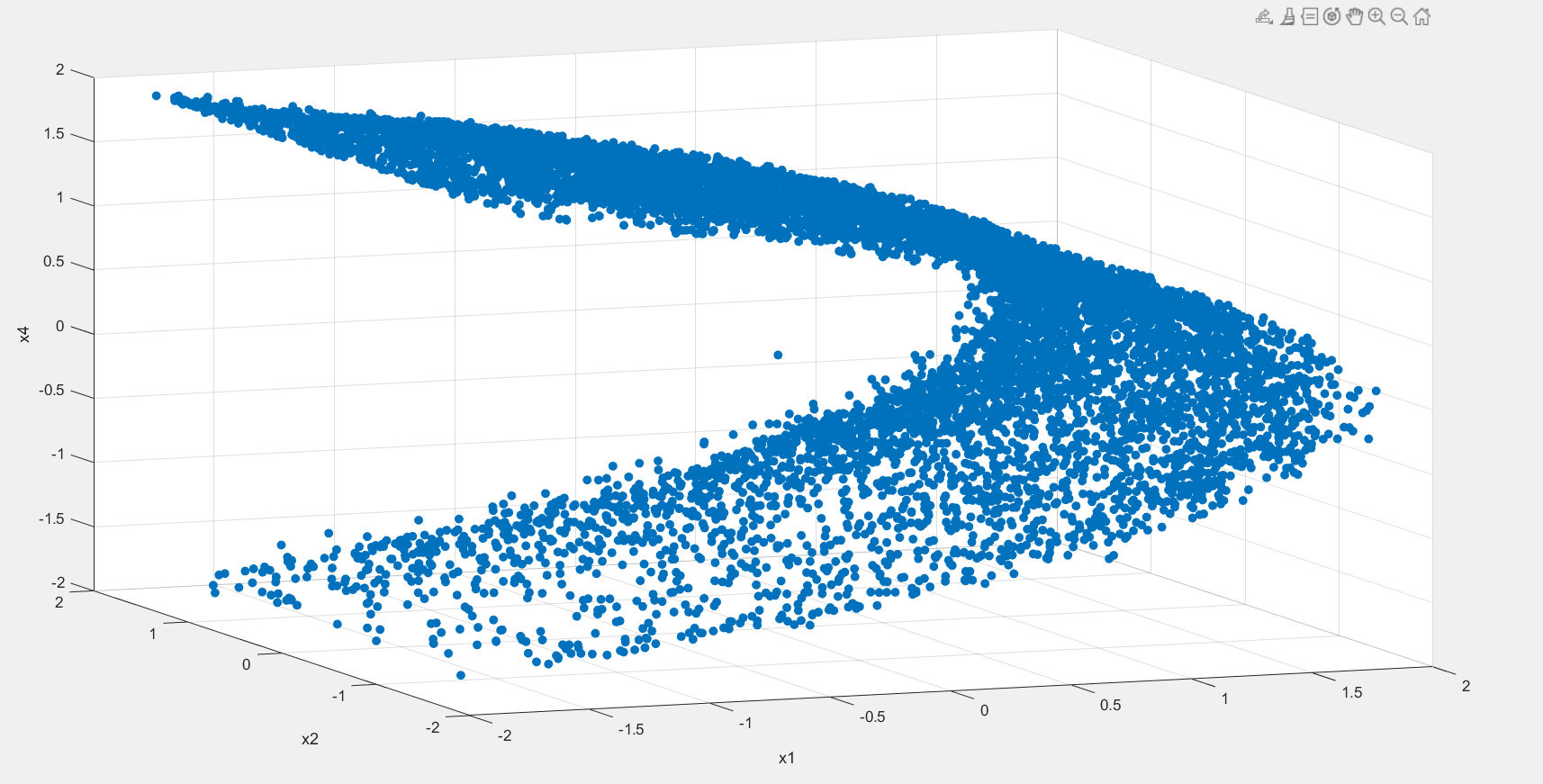}
         \label{fig:x1x2x4}
     \end{subfigure}
     \hfill
     \begin{subfigure}[b]{0.45\textwidth}
         \centering
         \includegraphics[width=\textwidth]{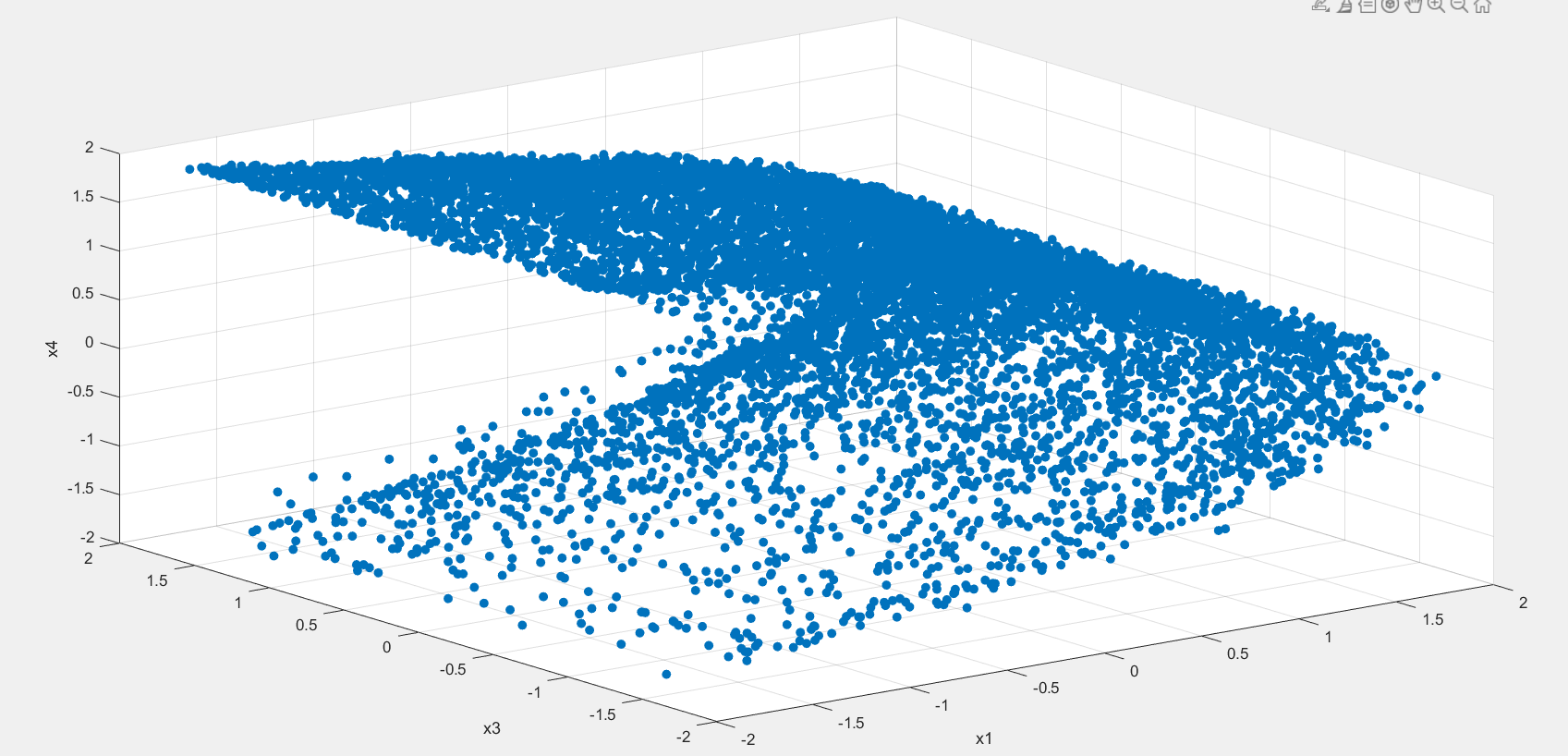}
         \label{fig:x1x3x4}
     \end{subfigure}
     \caption{State trajectories of the 4D hyperchaotic Henon map}
\end{figure}

For a system to exhibit hyperchaos, it requires a minimum of four dimensions. 
Due to this, we choose a 4D Henon map to scramble the transmitter data. 
A higher-order Henon map can generate hyperchaos in the cascade of scramblers without the need for generating newer maps.  
Hyperchaos occurs when a system has two positive, one null and one negative Lyapunov exponents. Due to the extra positive Lyapunov exponent, hyperchaotic systems exhibit more complex dynamical characteristics than chaotic systems.

\subsection{Chaotic Sequence Generator}
The transmission and receiver system is defined by Equation \ref{eq:henon_map} and \ref{eq:recv_system} respectively. 
 
\begin{equation}
  \label{eq:recv_system}
\begin{gathered}
y_1(k+1)  =a-y_3^2(k)+-by_4(k)
\\
y_2(k+1)  =y_1(k)
\\
y_3(k+1)   =y_2(k)
\\
y_4(k+1)   =y_3(k)
\end{gathered}
\end{equation}

Where, \(y_1(k)\), \(y_2(k)\), \(y_3(k)\), \(y_4(k)\) are the state variable of the receiver and \(a\), \(b\) are system parameters. Let \(Y_c =[y_1,y_2,y_3,y_4]^T\).
\(u(k)\) is defined as the sliding mode controller developed for projective synchronization (PS) of the two systems. 

The controller brings $\lim_{x \to \infty} $$|e(k)|$$ \ =0 $,  where \(\alpha\) is the scaling factor for synchronization. The error is modeled according to the following equations.
\begin{equation}
  e(k) = y(k)- \alpha(k)  
\end{equation}

The error for each dimension is calculated as shown in Equation \ref{eq:error}. 

\begin{equation}
  \label{eq:error}
\begin{gathered}
e_1(k+1)  =(1-\alpha )a + \alpha x_3^2(k)-y_3^2(k)-be4(k)
\\
e_2(k+1)  =e_1(k)
\\
e_3(k+1)   =e_2(k)
\\
e_4(k+1)   =e_3(k)
\end{gathered}
\end{equation}

\subsection{Synchronization of the 4D Henon Maps}
The controller used for the synchronization is a sliding mode controller. 
A sliding mode controller is used as it is highly scalable with increasing dimensions, and the sliding surface is derived from the Lyapunov exponents of the Henon maps.

The sliding surface of the controller is defined by Equation \ref{eq:sliding_surface}.

\begin{equation}
    \label{eq:sliding_surface}
    S(k)=e_1(k)+c_1 e_2(k)+c_2 e_3(k)+c_3 e_4(k)
\end{equation}

Where, \(c_1\), \(c_1\) and \(c_1\) are the design parameters given by the root of  \(P(s)=s^3+c_1s^2+c_2s+c2=0\).

From the sliding surface in Equation \ref{eq:sliding_surface}, the controller is derived as \cite{AbdulRahman2019SecureSystems}

\begin{multline}
\label{eq:controller}
u(k) = (1-qT)S(k) - \epsilon_1 (\lvert S(k) \rvert )^\beta sgn(S(k)) 
- \epsilon_2T(\lvert S(k) \rvert  )^\gamma sgn(S(k))\\ - c_1e_1(k) 
- c_2e_2(k)
- c_3e_3(k)+ be_4(k) - \alpha x_3^2(k)+ y_3^2(k) - a + \alpha    
\end{multline}

Where \(T>0, \epsilon_1>0, \epsilon_2>0, q>0, (1-qT)>0, 0<\beta<1\) and \(\gamma>1\). 
The controller ensures as $ k \to +\infty$, the errors in all the dimensions \(e_1, e_2, e_3, e_4\) will converge to zero.

\section{System Design}
\label{section:System Design}

\begin{figure*}[h]
\begin{center}
\includegraphics[width=\textwidth]{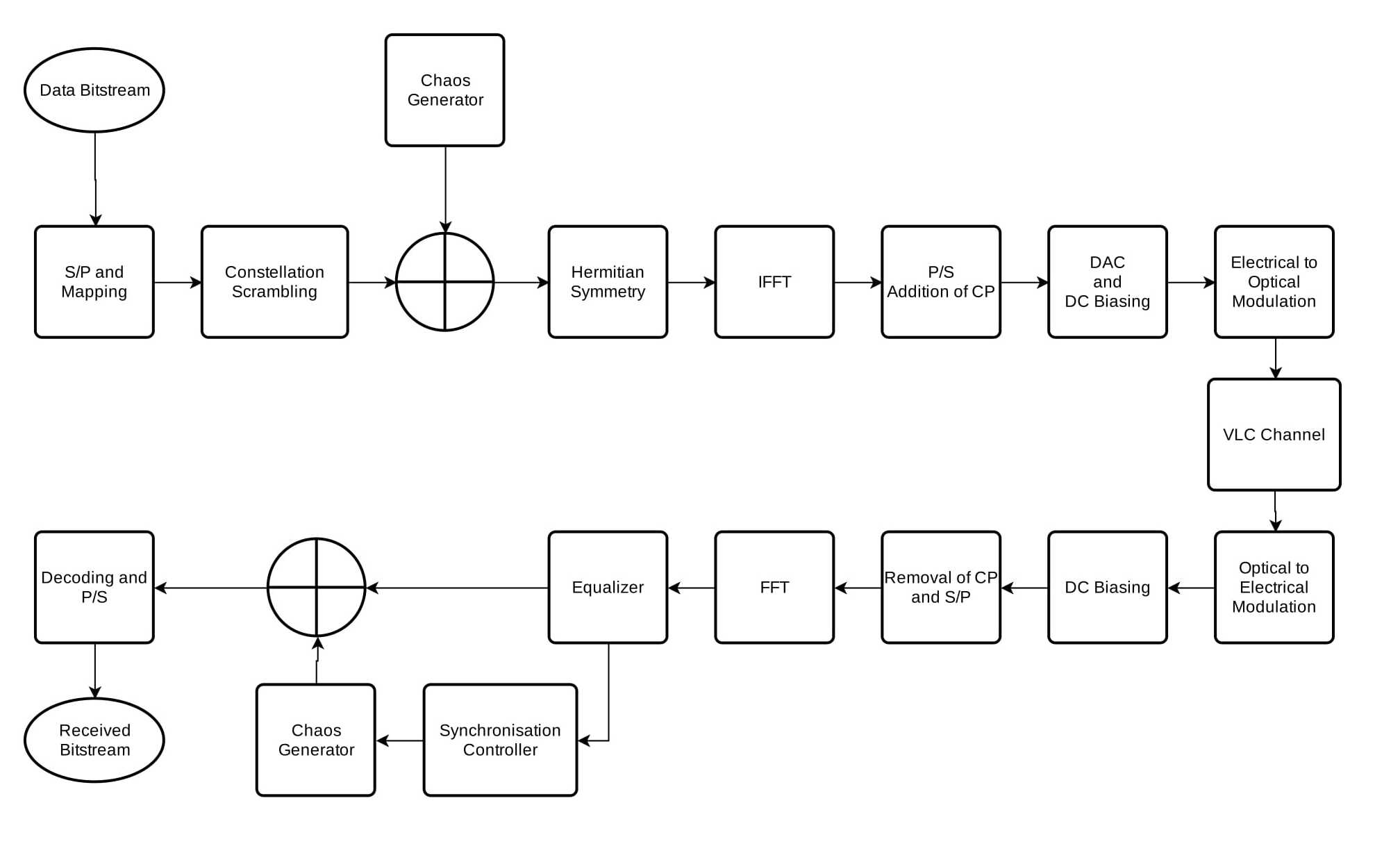}
\end{center}
 \caption{System schematics}
 \label{fig:Schematic}
\end{figure*}

The system is designed to be a DCO-OFDM modulation using IM/DD. A brief overview of the design of the model is presented in Figure \ref{fig:Schematic}. 

The data is initially converted from serial to parallel and mapped using the PSK/QAM sequence.
The M-QAM mapped data is given by \(X=[X_1, X_2 ... X_m]^T\), which is then scrambled using the cascade of scramblers with the help of the 4D Henon map generator. The data is then set to hermitian symmetry  \(X'=[0,X'_1 ... X'_m,0,X'_1 ... X'_m]^T\). 
The data is transformed to hermitian symmetry to get real values when the data goes through Inverse Fast Fourier Transform (IFFT), as LEDs cannot transmit imaginary data. The data can be represented as

\begin{equation}
\label{eq:data_model}
S_{i,j}=\frac {1}{N_{d}}\sum \limits _{n=1}^{N_{d} }{{ {{X'}}}_{i,n}e^{\left ({J\frac {2\pi }{N}(n\mathrm {-1)(}j\mathrm {-1)} }\right)}}
\end{equation}

Where \(S_{i,j}\) and \(X_{i,j}\), are the \(j^{th}\) and \(k^{th}\) samples of \(S_i\) and \(X_i\) , respectively.
The data is transformed from parallel to serial, and a cyclic prefix (CP) is added.\
 
Then it is transformed from analog to digital, and DC biasing is added, which is essential as LEDs cannot transmit negative voltage. The data is finally converted from electrical to an optical signal and transmitted via a visible light communication channel described in Section \ref{section:Communication Channel}.

The incoming data is first converted from an optical signal to an electrical signal using a photodiode. This signal is then filtered at the receiver side, and gain is added before converting it to a digital signal from analog.
After the CP is removed, the signal goes through a Fast Fourier Transform (FFT), and hermitian symmetry is removed.
We use the Least-Square Method for channel estimation.
This then works as an input to the sliding mode controller and gives the synchronization constant,  \(u(k)\). This synchronization constant is fed to the receiver's chaotic generator, which helps synchronize the two hyperchaotic maps.
The data is then de-scrambled using the synchronized 4D Henon map, and the mapped data is finally converted from parallel to serial.

\subsection{Constellation Scrambling}

Scrambling is a common technique in encryption. There have been many approaches to scrambling. We propose a technique for constellation symbol scrambling using the 4D hyperchaotic Henon map. 

The scrambler is defined as 

\begin{equation}
  \label{eq:scrambler}
\begin{gathered}
    X_{real} = real(X)*sign(x_a)\\
    X_{imaginary} = imaginary(X)*sign(x_b)\\
    X = X_{real} +X_{imaginary}
\end{gathered}
\end{equation}

Where \(X\) is the mapped constellation matrix of the system and \(x_{a}\), \(x_{b}\) are the dimensions of the Henon map. 

The hyperchaos generator gives a random sequence of numbers of the range \((-2, 2)\). Thus, this scrambling system is entirely random without the actual key to the synchronization sliding mode controller.

Since we use a 4D Henon map, we can improve the security by implementing a cascade of scramblers. This provides more security than single-layered scramblers by encoding the data into higher dimensions. 
With the given four dimensions of the map, we can create a cascade of six scramblers using different mapping dimensions.


\subsection{Complexity}
In the proposed method, most complexity is due to the chaotic generator, IFFT and the channel scrambling. The IFFT requires \(((N_d)/2)\log_2 N_d\) complex multiplication.  The computational complexity of the chaotic generating sequence with N is of the order of O(N). The last part contributing to the system's complexity is the scrambling operations, which is of the order of O(N). In the proposed system of 16QAM OFDM, we take the length of the scrambled digitally modulated symbol sequence needed is \(N /4\). 

Thus the total complexity for six scrambler in cascade can be written as.
\begin{equation}
    O(N\log_2 N) +O(N) +6*2*O(N/4)\\
\end{equation}

 And, total complexity for a single scrambler can be represented as
 \begin{equation}
    O(N\log_2 N) +O(N) +2*O(N/4)\\
\end{equation}

\section{Communication Channel}
\label{section:Communication Channel}

\subsection{LOS Impulse Response}
\begin{figure}[h]
    \centering
   \includegraphics[height=4cm]{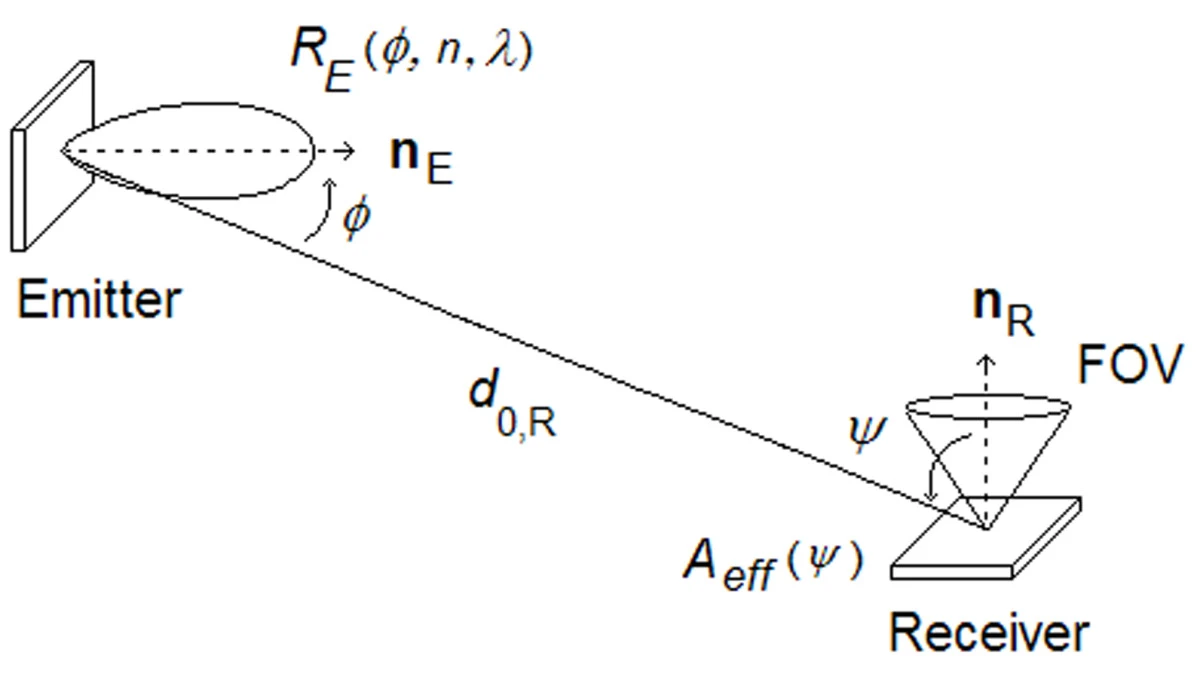}
    \caption{Geometry of emitter and receiver without reflectors}
    \label{fig:tx_rx_geometry}
\end{figure}

The Line of Sight (LOS) impulse response, given an emitter E and receiver R with distance \(d_{0,R}\) as shown in Figure \ref{fig:tx_rx_geometry} is given by \cite{Rodriguez2013}

\begin{equation}
    \label{eq:}
    h^{\left(0\right)}\left(t;E,R,\lambda\right)=\frac1{\left(d_{0,R}\right)^2}R_E\left(\varphi,n,\lambda\right)A_{\mathrm{eff}}\left(\psi\right)\delta\left(t-\frac{d_{0,R}}c\right)
\end{equation}

where \(R_E\left(\varphi,n,\lambda\right)\) represents the generalized Lambertian model used to approximate the radiation pattern of the emitter, c is the speed of light and \(A_{eff}\) is the effective signal-collection area of the receiver, which is given by

\begin{equation}
    A_{\mathrm{eff}}\left(\psi\right)=A_R\operatorname{cos}\left(\psi\right)\hspace{0.25em}\mathrm{rect}\left(\frac\psi{\mathrm{FOV}}\right)
\end{equation}

The emitter is generally modelled using a Lambertian radiation pattern for each wavelength and has axial symmetry (independent of $\gamma$). Thus the receiving power can be represented as

\begin{equation}
    \begin{array}{c}R_E\left(\varphi,n,\lambda\right)=\frac{n+1}{2\pi}P_E\left(\lambda\right)\operatorname{cos}^n\left(\varphi\right),\\\hspace{7em}-\frac\pi2\leq\varphi\leq\frac\pi2,\hspace{1em}0\leq\gamma\leq2\pi\end{array}
\end{equation}

Where \(n\) is the mode number of the radiation lobe, which specifies the directionality of the emitter, integrating \(P_E(\lambda)\) over the emitter wavelength interval yields the nominal power \(P_E\) emitted by the emitter.

\subsection{Noise Model}
A Free-Space Optical (FSO) communication channel with the help of LEDs is used. The received signal can be represented as 

\begin{equation}
    \label{eq:received_signal}
    I_p(t)=RP_t(T)\otimes h(t) + n(t)
\end{equation}

Where \(R\) is the photodiode responsivity, \(P_t(t)\) the instantaneous transmitted power, \(h(t)\) is the channel impulse response, and \(n(t)\) is the variance of Total noise for the VLC System which is represented as the variance of the sum of shot noise and thermal noise \cite{Komine2004}.

\begin{equation}
    N_{total} = \sqrt{N_{thermal}^2 +N_{shot}^2}
\end{equation}

Where, \(N_{total}\) is the total noise in the system, \(N_{shot}\) is the shot noise, and \(N_{thermal}\) is the thermal noise.
 
Shot noise and Thermal noise is represented as 

\begin{equation}
    \label{eq:thermal_shot_noise}
    \begin{split}
        N_{shot}^2 = 2q(P_r +P_n)B \\
        N_{thermal}^2 = \frac{4KT}{BR_f}
    \end{split}
\end{equation}

Where \(q\) is the charge on an electron, \(P_r\) is the received noise power, \(P_n\) is the received optical power, and \(B\) is the detector bandwidth. Moreover, \(K\) s the Boltzmann constant, \(T\) is the absolute temperature and \(R_f\) is the feedback resistance.

Thus, the received signal can be represented as
\begin{equation}
    Y(t) = TX(t)*H_{loss} + N_{total}(t)
\end{equation}

Where \(Y(t)\)  the received signal,\(X(t)\)  the transmitted signal, \(H_{loss}\) is the channel impulse response which is taken as a constant as, during the results, the data rate has not been changed. 

\subsection{Information Leakage}
Assuming that 0 and 1 have equal probability of getting transmitted,
the mutual information between the transmitted data X and the data Y recovered by eavesdroppers is.

\begin{equation}
     I_k(Y;TX) = H_k(Y) -H_k(Y \textbar TX)
    = 1 + p_k{\log_2 p_k} +(1-p_k){\log_2 (1-p_k)}    
\end{equation}

Where \(H_k(.)\) represents the entropy operation and \(p_k\) is the BER for each user of the proposed system.

\section{Simulation}
\label{section:Simulation}
In this section, we evaluate the performance of the proposed system and compare it against existing results. 
We use BER and Information Leakage to represent the increased security of the system. 
The proposed VLC system is simulated using Matlab and Simulink to show the performance achieved for various modulation schemes. The simulation parameters are summarised in Table \ref{table:simulation_parameters_fso}.

\begin{table}
\begin{center}
\begin{tabular} { c|c } 
 \hline
 \textbf{Notation}                 & \textbf{Definition}      \\ 
 \hline \hline
 Symbol Rate                         & \(1 Gbps\)           \\ 
 \hline
 Room Size                           & \(5*3*3\ m\)         \\
 \hline
 Transmitter Position                & \((0,0,3) m\)        \\
 \hline
 Receiver Position                   & \((2.5,2.5,0) m\)    \\
 \hline
 LED Power                           & \(3.2W\)             \\
 \hline
 Active Area                         & \(1cm\)              \\
 \hline
 Half Angle FOV                      & \(85\degree\)        \\
 \hline
 Semi-angle at half power            & \(60\degree\)        \\
 \hline
 Optical Filter Gain                 & \(1.0\)              \\
 \hline
\end{tabular}
\end{center}
\caption{Simulation parameters of VLC}
\label{table:simulation_parameters_fso}
\end{table}

\begin{figure}[h]
    \centering
   \includegraphics[height=4cm]{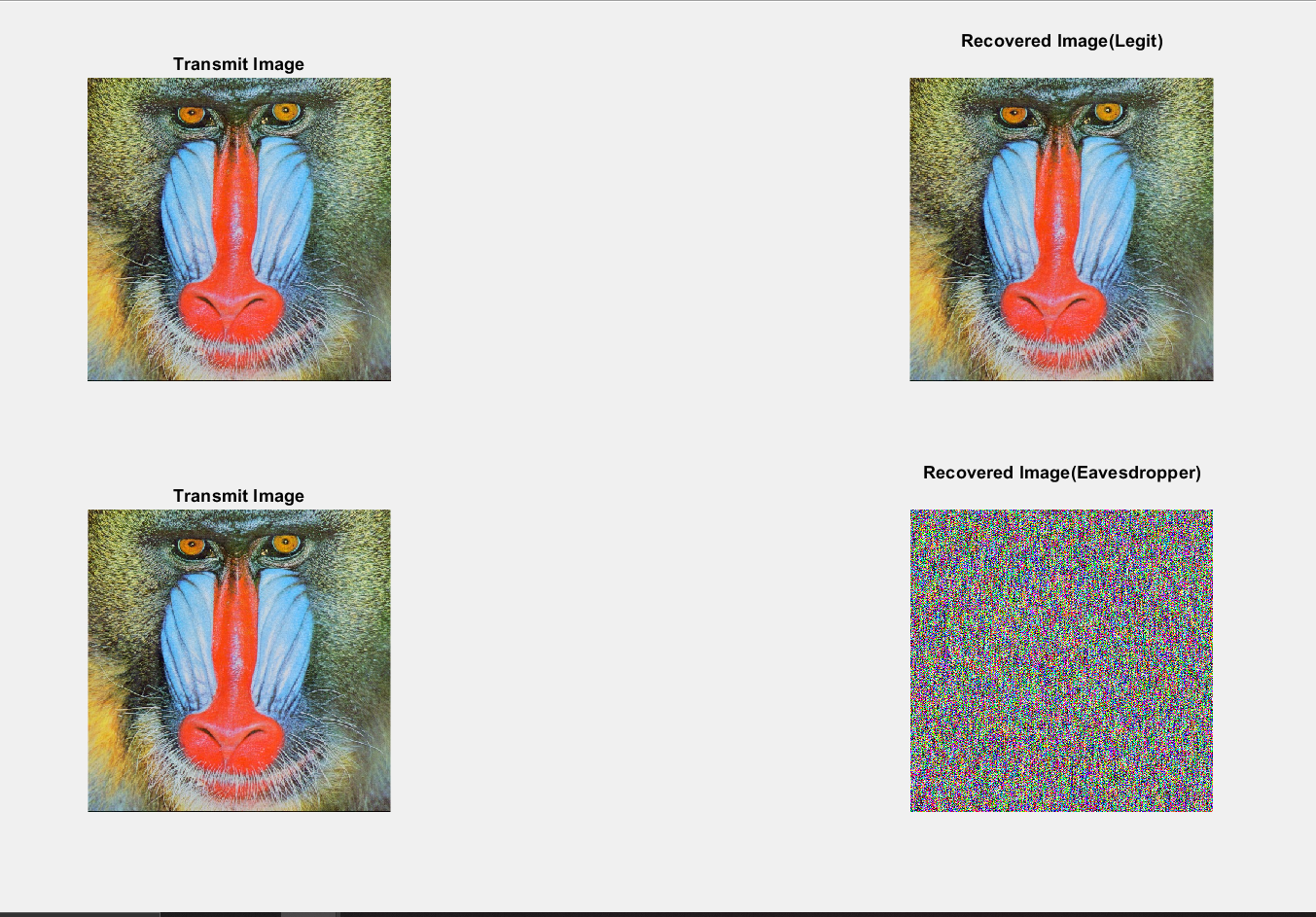}
    \caption{Transmitted and received photo for legitimate user and eavesdropper}
    \label{fig:ber_var_synchronization}
\end{figure}

The initial condition for the two 4D Henon maps are given by \(x_1(0)=0.1\), \(x_2(0)=-0.1\), \(x_3(0)=0.1\), \(x_4(0)=0.1\), \(y_1(0)=0.3\), \(y_2(0)=-0.1\), \(y_3(0)=0.2\), \(y_4(0)=0.1\), and parameters for the sliding mode controller are given by
\(T= 1\), \(\epsilon_1=0. 1\), \(\epsilon_2= 0. 1\), \(q= 0. 7\), \(\alpha = 0. 8\), \(\gamma = 1. 2\),
\(\beta = 0. 9\), \( c_1 =0.001\), \(c_2=0\), \(c_3=0\) where the scaling factor is \(\alpha =0.8\).

\begin{figure}[h]
    \centering
   \includegraphics[height=3.8cm]{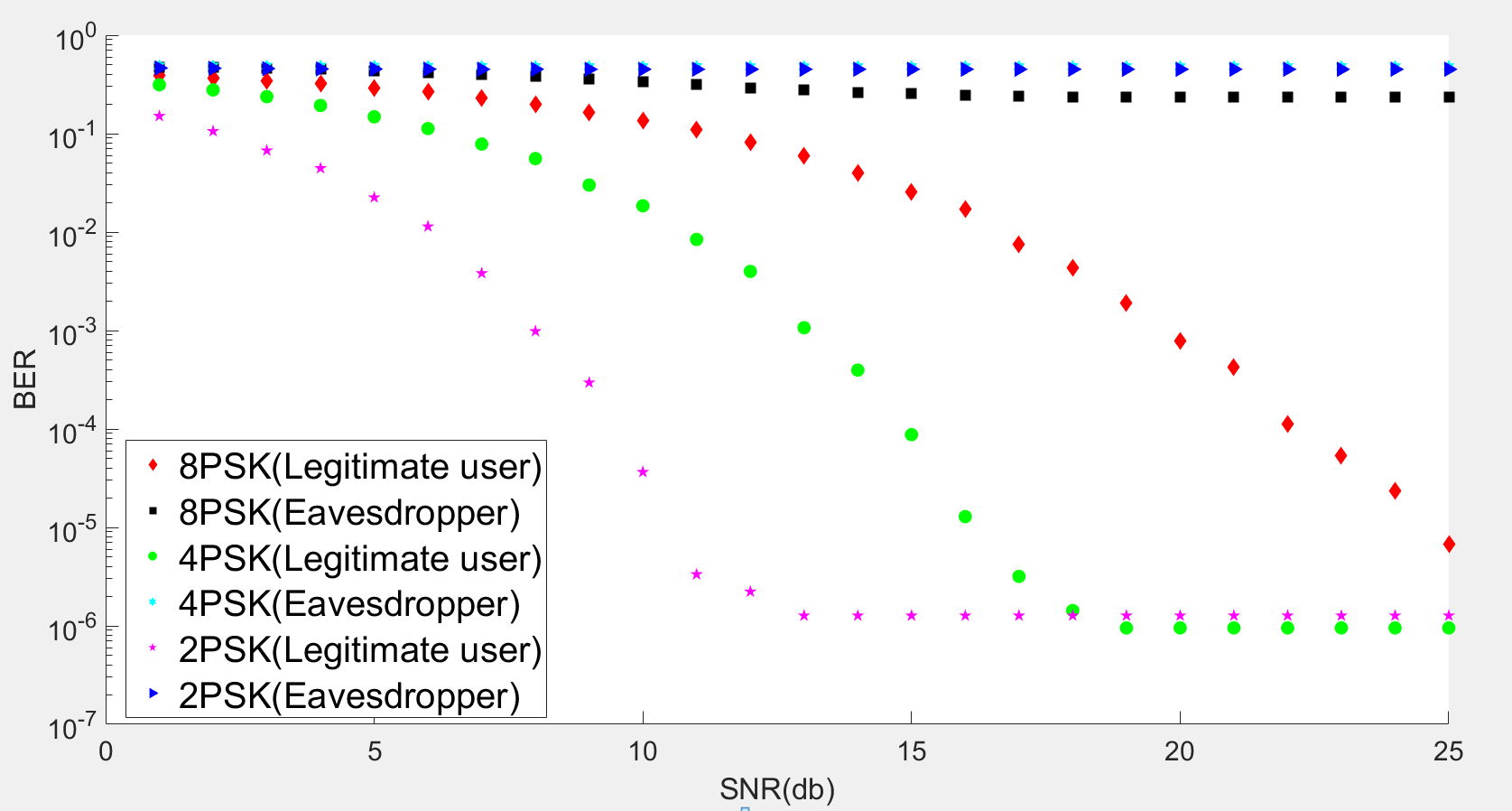}
    \caption{BER vs SNR of different modulation schemes for legitimate user and eavesdropper}
    \label{fig:ber}
\end{figure}

\begin{figure}[h]
    \centering
   \includegraphics[height=3.8cm]{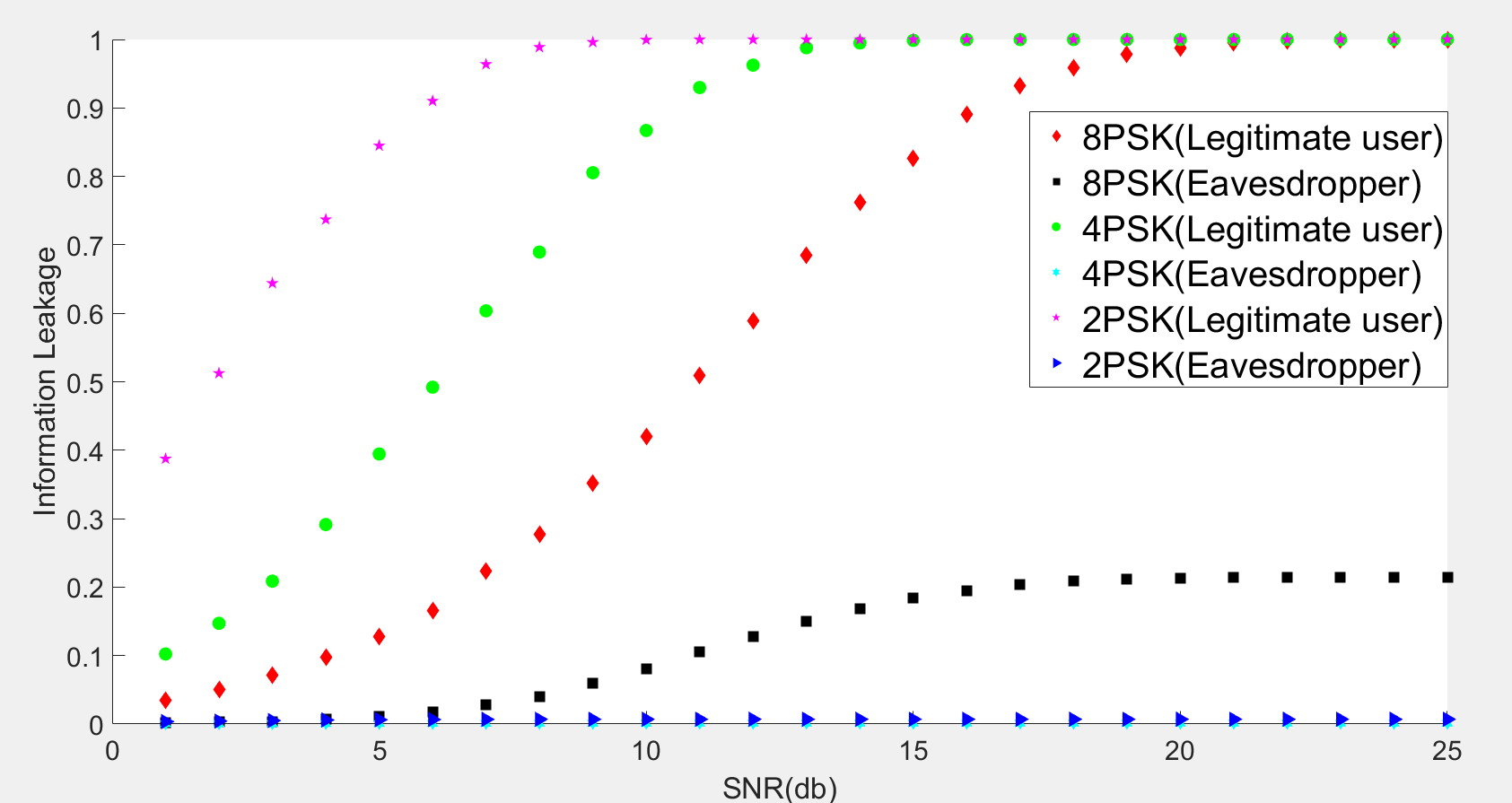}
    \caption{Information Leakage vs SNR}
    \label{fig:info_leak}
\end{figure}

\begin{figure}[h]
    \centering
   \includegraphics[height=3.8cm]{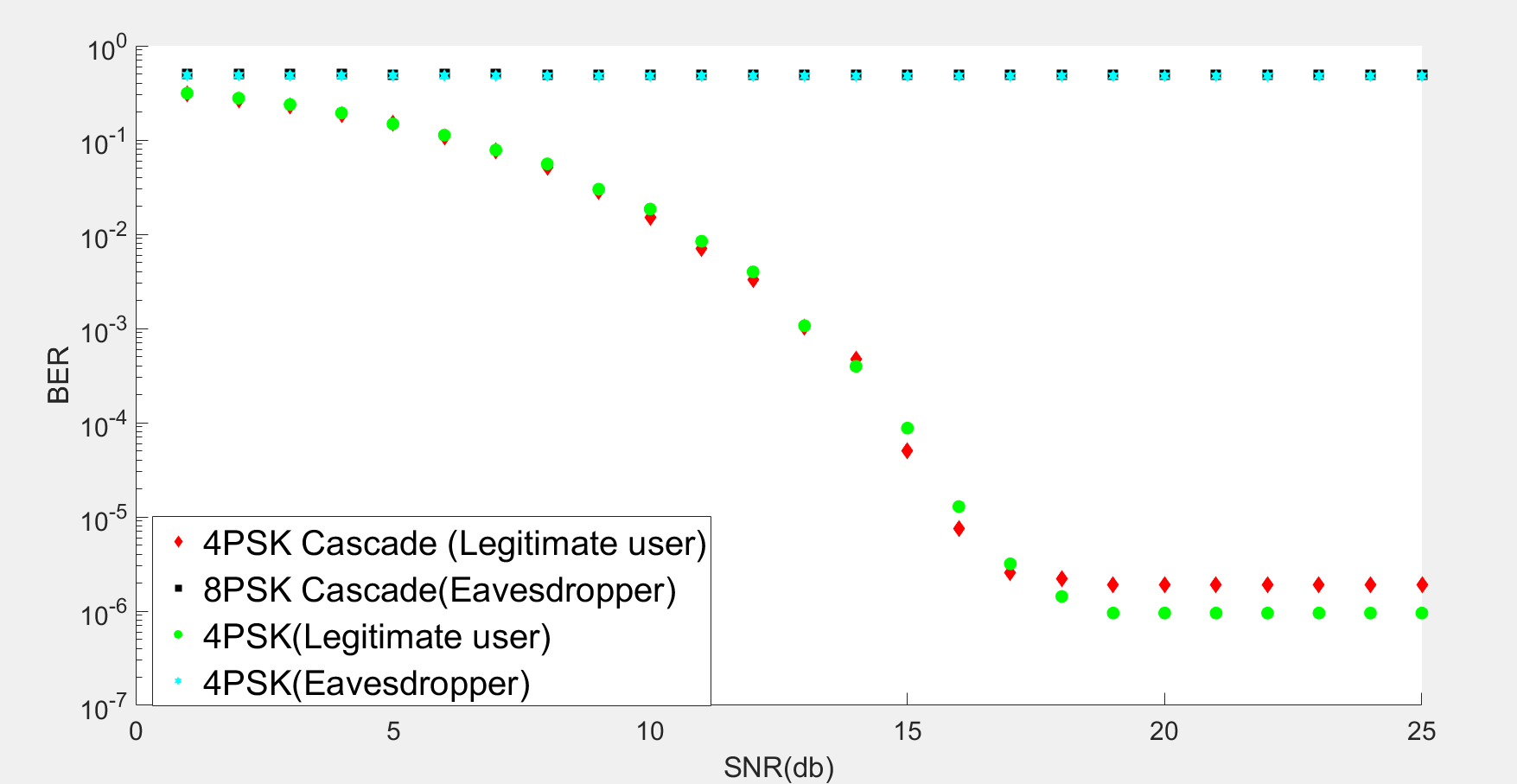}
    \caption{BER for single and cascading encryption schemes}
    \label{fig:single_vs_cascade_ber}
\end{figure}

The environment is set up as an empty room of size 5x5m and height 3m. 
A \((5,5,3)m\) empty room is used as the simulation environment. The transmitter is in the middle of the room on the ceiling \((0,0,3)m\). The receiver is on the ground near the end of the wall \((2.5,2.5,0)m\).

Figure \ref{fig:ber_var_synchronization} shows the transmitted image and the received image for a legitimate user as well as for eavesdropper. A clear image is shown for the legitimate users, whereas almost no information is received for the eavesdropper. We achieve a very low information leakage for the system of \(10^{-5}\), indicating better results compared to the existing system of around \(0.1\) \cite{Chen2016}.

Figure \ref{fig:ber} presents a relation between the BER and Signal-to-Noise Ratio (SNR) using  BPSK, QPSK and 8PSK modulation schemes for legitimate users and eavesdroppers to represent the security of the system in case of no key. 
Without the key, the BER of the system for all spatial modulation does not drop below \(0.49\), exclaiming the capacity of security. 
On the other hand, a relatively low BER can be seen for adequate SNR. We achieve a BER of \(10^{-6}\), which is similar to existing results with decreased complexity of the system, making it easier to implement on IoT devices \cite{XiaogeWu2016}. Moreover, the increased BER of \(0.49\) indicates better security than the present system.



 \begin{figure}[b]
     \centering
     \begin{subfigure}[b]{0.5\textwidth}
         \centering
         \includegraphics[width=\textwidth]{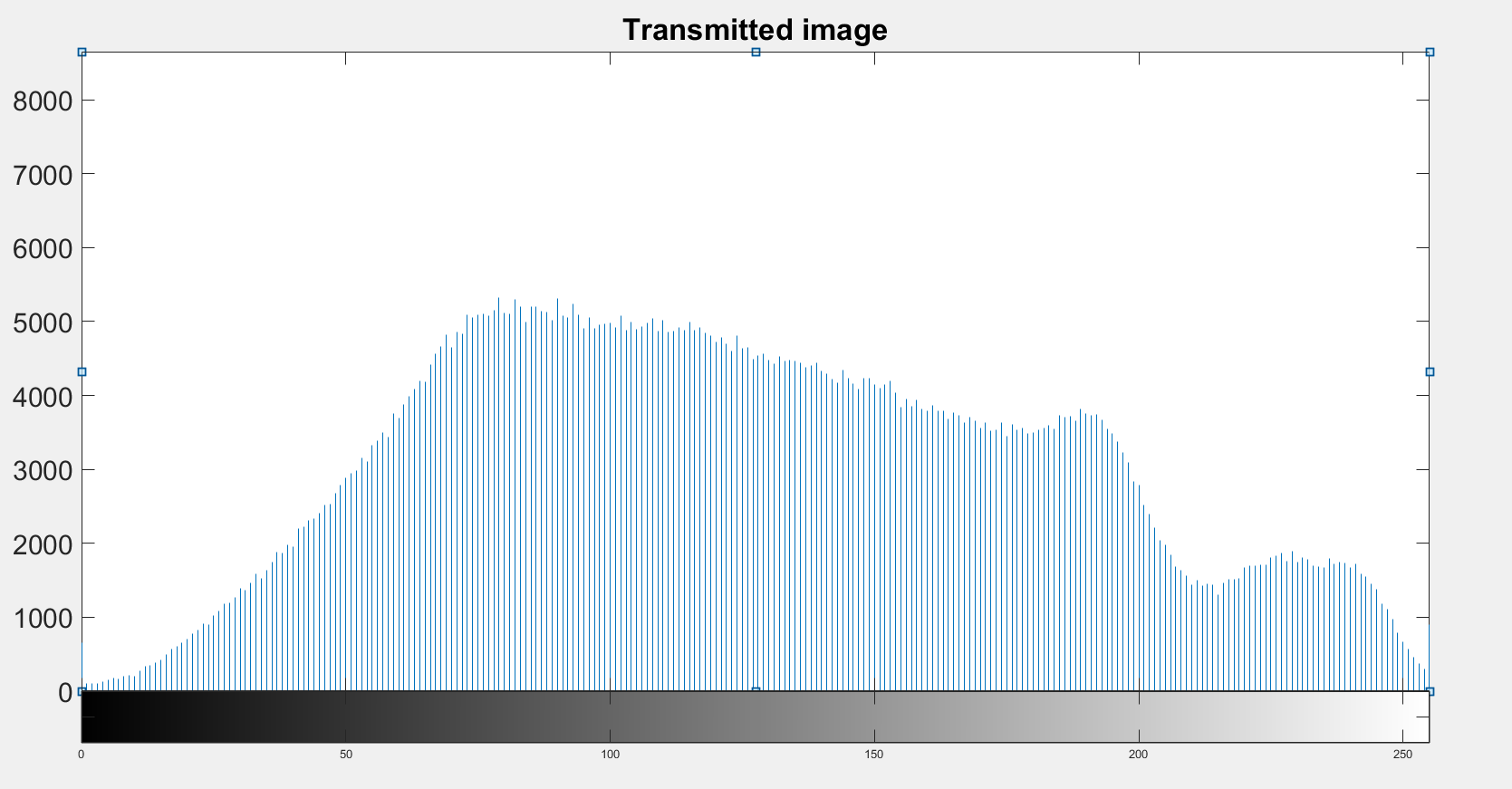}
         \caption{Histogram of transmitted image}
         \label{fig:y equals x}
     \end{subfigure}
     \hfill
     \begin{subfigure}[b]{0.5\textwidth}
         \centering
         \includegraphics[width=\textwidth]{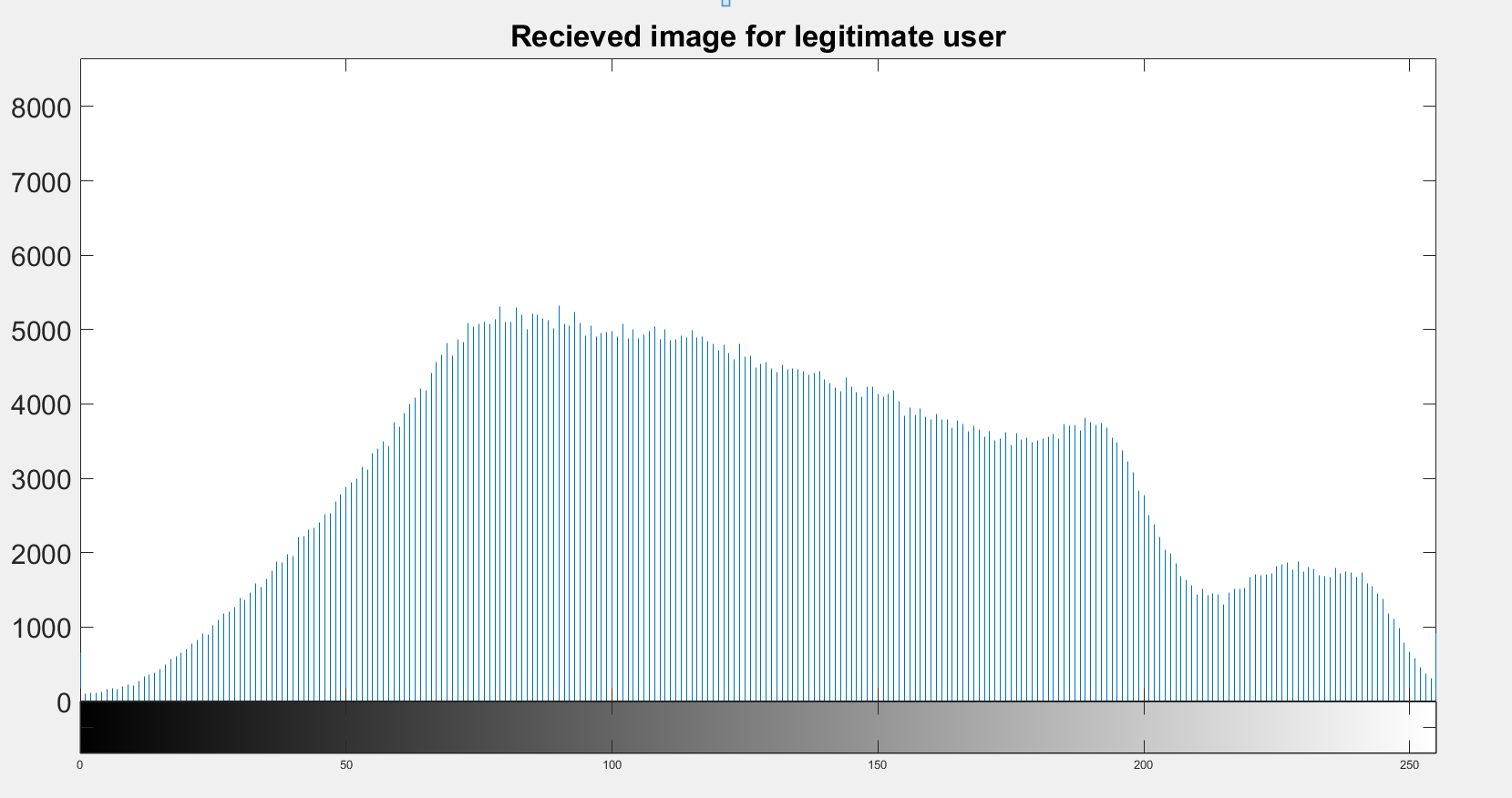}
         \caption{Histogram of Received Image for Legitimate user}
         \label{fig:three sin x}
     \end{subfigure}
     \hfill
     \begin{subfigure}[b]{0.5\textwidth}
         \centering
         \includegraphics[width=\textwidth]{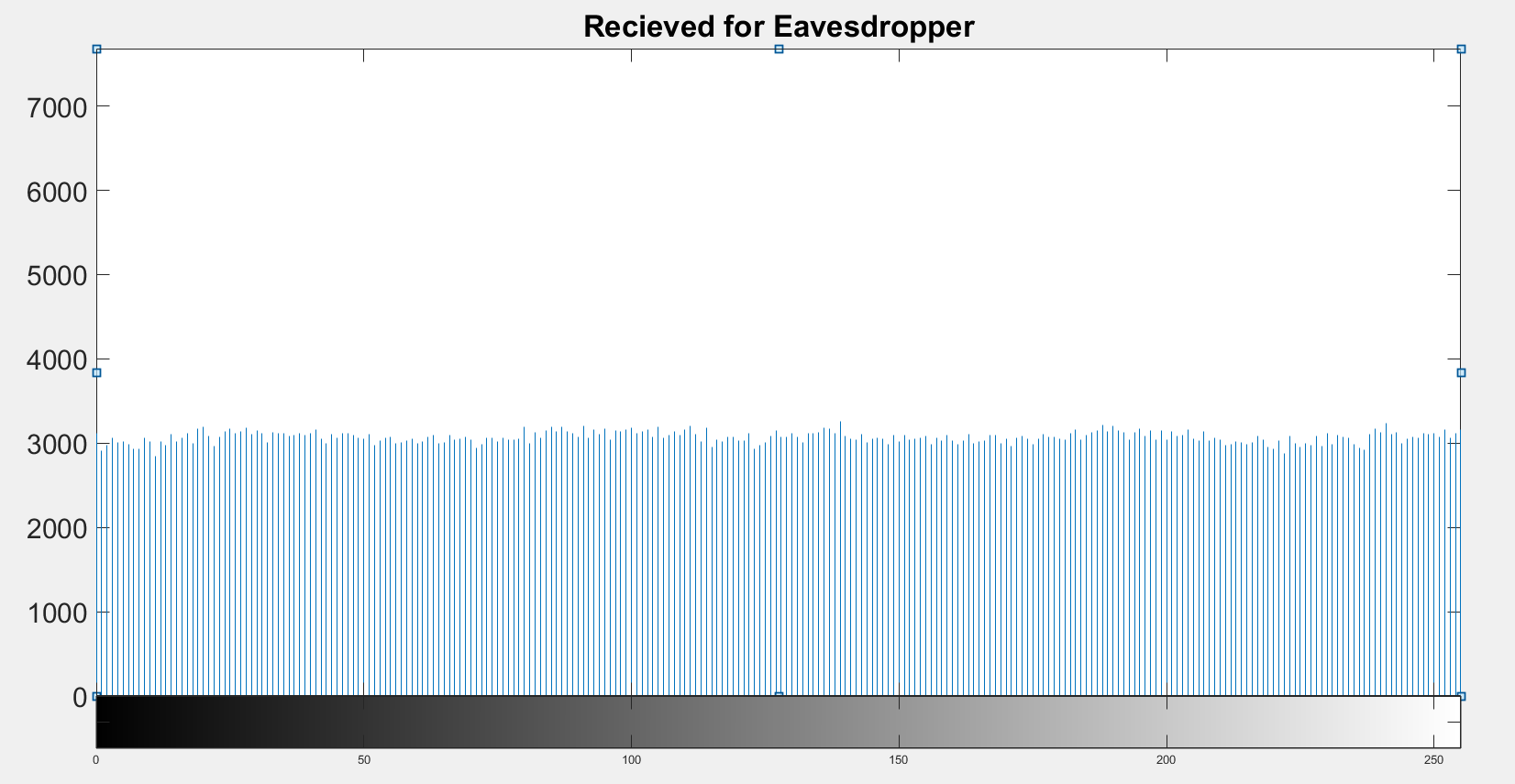}
         \caption{Histogram of Received Image for Eavesdropper}
         \label{fig:five over x}
     \end{subfigure}
        \caption{Histogram for transmitted and received image}
        \label{fig:hist}
\end{figure}

Figure \ref{fig:info_leak} demonstrates the Information Leakage of the proposed system at different SNR for the previously mentioned spatial modulation schemes, with and without the key. The graph indicates that the low-security leakage for eavesdroppers, not going above \(0.1\). Moreover, all legitimate user modulation schemes go to one below \(25 dB\) SNR.

\begin{figure}[h]
    \centering
   \includegraphics[height=3.8cm]{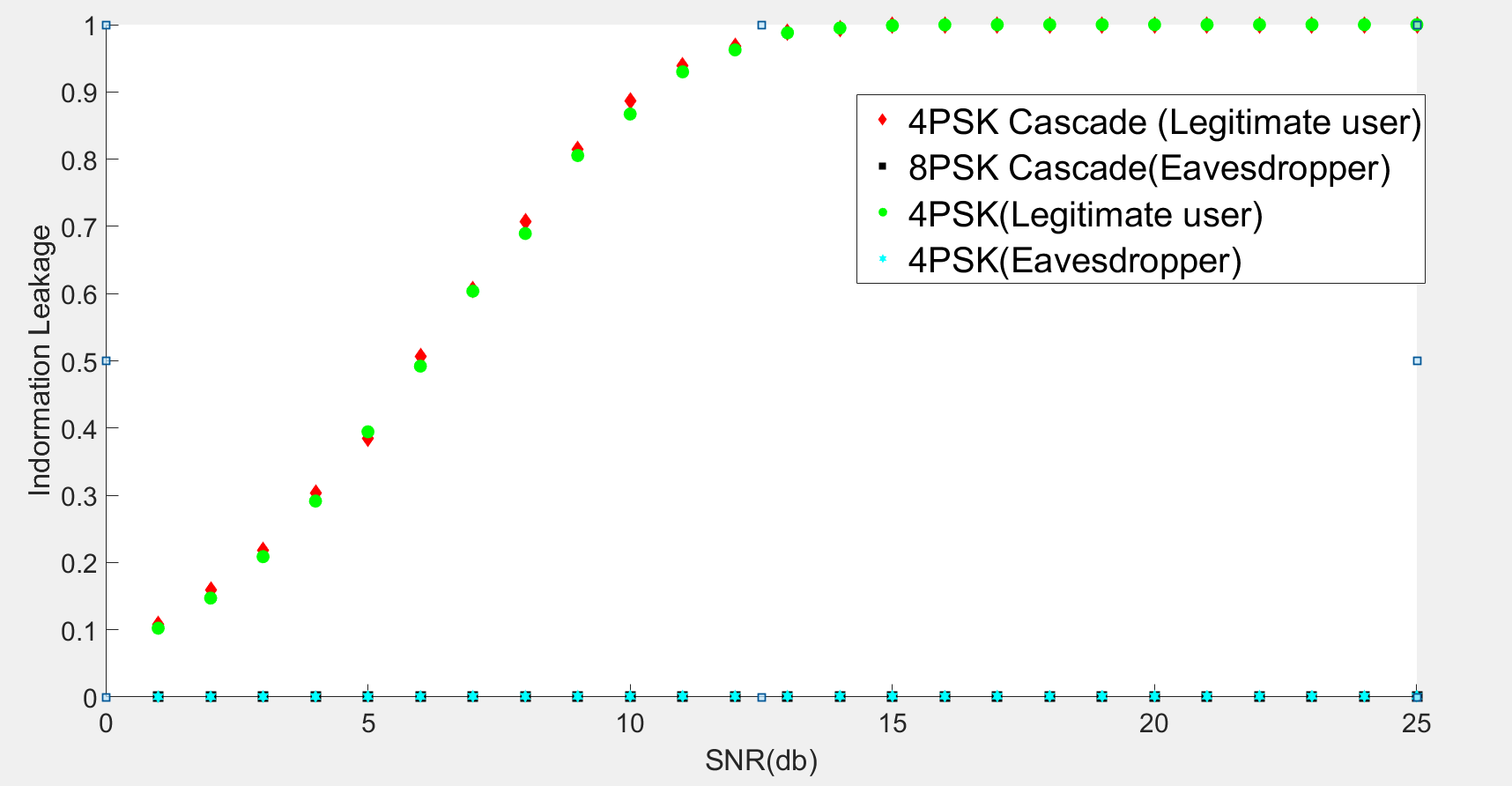}
    \caption{Information Leakage for single and cascading encryption schemes}
    \label{fig:si}
\end{figure}

Figure \ref{fig:single_vs_cascade_ber} compares the BER achieved by the two encryption schemes proposed. The cascaded encryption consists of six scramblers, where all the dimensions of the Henon map are used more effectively to scramble the constellations than the single-layered encryption.
Cascaded systems provide more security than single-layered systems. However, a higher BER is observed for both legitimate users and eavesdroppers than single-layered systems. The complexity of a cascaded system is also relatively higher.


Figure \ref{fig:hist} demonstrates the transmitted and the received histograms of the image for the legitimate user and the eavesdropper. The histogram for the legitimate users is near perfect to the transmitted image, whereas for the eavesdropper, it is not recognizable. On testing the transmitted image and received using the Two-sample Kolmogorov-Smirnov test, the \(p\)-value for legitimate users came out to be one. On the other hand, it came out to be  \(1.5252e-14\) for an eavesdropper. The system was tested for QPSK with cascade encryption and SNR \(25db\).

\section{Conclusion}
\label{section:Conclusion}
A system design for increasing the physical layer security has been proposed for visible light communication-based systems. The system gives a versatile and less complex approach where one can choose the number of scramblers depending on the SNR for optimal transmission. The BER was calculated for different modulation schemes for legitimate users and eavesdroppers, and the best results were obtained for 8PSK modulation. 
The system achieves a BER of \(10^{-6}\) for and SNR of \(25db\) for the 8PSK modulation scheme, showing the system's usability. Moreover, very low BER and information leakage are seen for an eavesdropper, showing the security prowess of the system.

Future work will be addressed towards doing an experimental analysis for the proposed system and finding new chaotic modulation schemes to enhance the security and usability of the proposed system. 

\section{Statements and Declarations}
All authors certify that they have no affiliations with or involvement in any organization or entity with any financial interest or non-financial interest in the subject matter or materials discussed in this manuscript.

\backmatter

\bmhead{Acknowledgments}

We thank our guide, Dr Ujjwal Verma, for his constant support and guidance. We would also like to thank Mars Rover Manipal for providing us with the necessary resources to complete the research.


\bibliography{sn-bibliography}


\begin{thebibliography}{32}
\ifx \bisbn   \undefined \def \bisbn  #1{ISBN #1}\fi
\ifx \binits  \undefined \def \binits#1{#1}\fi
\ifx \bauthor  \undefined \def \bauthor#1{#1}\fi
\ifx \batitle  \undefined \def \batitle#1{#1}\fi
\ifx \bjtitle  \undefined \def \bjtitle#1{#1}\fi
\ifx \bvolume  \undefined \def \bvolume#1{\textbf{#1}}\fi
\ifx \byear  \undefined \def \byear#1{#1}\fi
\ifx \bissue  \undefined \def \bissue#1{#1}\fi
\ifx \bfpage  \undefined \def \bfpage#1{#1}\fi
\ifx \blpage  \undefined \def \blpage #1{#1}\fi
\ifx \burl  \undefined \def \burl#1{\textsf{#1}}\fi
\ifx \doiurl  \undefined \def \doiurl#1{\url{https://doi.org/#1}}\fi
\ifx \betal  \undefined \def \betal{\textit{et al.}}\fi
\ifx \binstitute  \undefined \def \binstitute#1{#1}\fi
\ifx \binstitutionaled  \undefined \def \binstitutionaled#1{#1}\fi
\ifx \bctitle  \undefined \def \bctitle#1{#1}\fi
\ifx \beditor  \undefined \def \beditor#1{#1}\fi
\ifx \bpublisher  \undefined \def \bpublisher#1{#1}\fi
\ifx \bbtitle  \undefined \def \bbtitle#1{#1}\fi
\ifx \bedition  \undefined \def \bedition#1{#1}\fi
\ifx \bseriesno  \undefined \def \bseriesno#1{#1}\fi
\ifx \blocation  \undefined \def \blocation#1{#1}\fi
\ifx \bsertitle  \undefined \def \bsertitle#1{#1}\fi
\ifx \bsnm \undefined \def \bsnm#1{#1}\fi
\ifx \bsuffix \undefined \def \bsuffix#1{#1}\fi
\ifx \bparticle \undefined \def \bparticle#1{#1}\fi
\ifx \barticle \undefined \def \barticle#1{#1}\fi
\bibcommenthead
\ifx \bconfdate \undefined \def \bconfdate #1{#1}\fi
\ifx \botherref \undefined \def \botherref #1{#1}\fi
\ifx \url \undefined \def \url#1{\textsf{#1}}\fi
\ifx \bchapter \undefined \def \bchapter#1{#1}\fi
\ifx \bbook \undefined \def \bbook#1{#1}\fi
\ifx \bcomment \undefined \def \bcomment#1{#1}\fi
\ifx \oauthor \undefined \def \oauthor#1{#1}\fi
\ifx \citeauthoryear \undefined \def \citeauthoryear#1{#1}\fi
\ifx \endbibitem  \undefined \def \endbibitem {}\fi
\ifx \bconflocation  \undefined \def \bconflocation#1{#1}\fi
\ifx \arxivurl  \undefined \def \arxivurl#1{\textsf{#1}}\fi
\csname PreBibitemsHook\endcsname

\bibitem{Blinowski2019SecuritySurvey}
\begin{barticle}
\bauthor{\bsnm{Blinowski}, \binits{G.}}:
\batitle{{Security of Visible Light Communication systems—A survey}}.
\bjtitle{Physical Communication}
\bvolume{34},
\bfpage{246}--\blpage{260}
(\byear{2019}).
\doiurl{10.1016/j.phycom.2019.04.003}
\end{barticle}
\endbibitem

\bibitem{Lu2015High-SecurityShifting}
\begin{barticle}
\bauthor{\bsnm{Lu}, \binits{H.}},
\bauthor{\bsnm{Zhang}, \binits{L.}},
\bauthor{\bsnm{Jiang}, \binits{M.}},
\bauthor{\bsnm{Wu}, \binits{Z.}}:
\batitle{{High-Security Chaotic Cognitive Radio System with Subcarrier
  Shifting}}.
\bjtitle{IEEE Communications Letters}
\bvolume{19}(\bissue{10}),
\bfpage{1726}--\blpage{1729}
(\byear{2015}).
\doiurl{10.1109/LCOMM.2015.2464214}
\end{barticle}
\endbibitem

\bibitem{Classen2016OpportunitiesLayer}
\begin{botherref}
\oauthor{\bsnm{Classen}, \binits{J.}},
\oauthor{\bsnm{Steinmetzer}, \binits{D.}},
\oauthor{\bsnm{Hollick}, \binits{M.}}:
Opportunities and pitfalls in securing visible light communication on the
  physical layer,
19--24
(2016).
\doiurl{10.1145/2981548.2981551}
\end{botherref}
\endbibitem

\bibitem{Elgala2011}
\begin{botherref}
\oauthor{\bsnm{Elgala}, \binits{H.}},
\oauthor{\bsnm{Mesleh}, \binits{R.}},
\oauthor{\bsnm{Haas}, \binits{H.}}:
Indoor optical wireless communication: Potential and state-of-the-art.
IEEE Communications Magazine
\textbf{49}
(2011).
\doiurl{10.1109/MCOM.2011.6011734}
\end{botherref}
\endbibitem

\bibitem{Elgala2009}
\begin{botherref}
\oauthor{\bsnm{Elgala}, \binits{H.}},
\oauthor{\bsnm{Mesleh}, \binits{R.}},
\oauthor{\bsnm{Haas}, \binits{H.}}:
Indoor broadcasting via white leds and ofdm.
IEEE Transactions on Consumer Electronics
\textbf{55}
(2009).
\doiurl{10.1109/TCE.2009.5277966}
\end{botherref}
\endbibitem

\bibitem{Kumar2010}
\begin{botherref}
\oauthor{\bsnm{Kumar}, \binits{N.}},
\oauthor{\bsnm{Lourenço}, \binits{N.R.}}:
Led-based visible light communication system: A brief survey and investigation.
Journal of Engineering and Applied Sciences
\textbf{5}
(2010).
\doiurl{10.3923/jeasci.2010.296.307}
\end{botherref}
\endbibitem

\bibitem{Sevincer2013}
\begin{botherref}
\oauthor{\bsnm{Sevincer}, \binits{A.}},
\oauthor{\bsnm{Bhattarai}, \binits{A.}},
\oauthor{\bsnm{Bilgi}, \binits{M.}},
\oauthor{\bsnm{Yuksel}, \binits{M.}},
\oauthor{\bsnm{Pala}, \binits{N.}}:
Lightnets: Smart lighting and mobile optical wireless networks - a survey.
IEEE Communications Surveys and Tutorials
\textbf{15}
(2013).
\doiurl{10.1109/SURV.2013.032713.00150}
\end{botherref}
\endbibitem

\bibitem{Wu2014}
\begin{botherref}
\oauthor{\bsnm{Wu}, \binits{S.}},
\oauthor{\bsnm{Wang}, \binits{H.}},
\oauthor{\bsnm{Youn}, \binits{C.H.}}:
Visible light communications for 5g wireless networking systems: From fixed to
  mobile communications.
IEEE Network
\textbf{28}
(2014).
\doiurl{10.1109/MNET.2014.6963803}
\end{botherref}
\endbibitem

\bibitem{Karunatilaka2015}
\begin{botherref}
\oauthor{\bsnm{Karunatilaka}, \binits{D.}},
\oauthor{\bsnm{Zafar}, \binits{F.}},
\oauthor{\bsnm{Kalavally}, \binits{V.}},
\oauthor{\bsnm{Parthiban}, \binits{R.}}:
Led based indoor visible light communications: State of the art.
IEEE Communications Surveys and Tutorials
\textbf{17}
(2015).
\doiurl{10.1109/COMST.2015.2417576}
\end{botherref}
\endbibitem

\bibitem{Qiu2016}
\begin{botherref}
\oauthor{\bsnm{Qiu}, \binits{Y.}},
\oauthor{\bsnm{Chen}, \binits{H.H.}},
\oauthor{\bsnm{Meng}, \binits{W.X.}}:
Channel modeling for visible light communications—a survey.
Wireless Communications and Mobile Computing
\textbf{16}
(2016).
\doiurl{10.1002/wcm.2665}
\end{botherref}
\endbibitem

\bibitem{Sindhubala2016}
\begin{botherref}
\oauthor{\bsnm{Sindhubala}, \binits{K.}},
\oauthor{\bsnm{Vijayalakshmi}, \binits{B.}}:
Survey on noise sources and restrain techniques in visible-light communication.
Light and Engineering
\textbf{24}
(2016)
\end{botherref}
\endbibitem

\bibitem{Zhuang2018}
\begin{botherref}
\oauthor{\bsnm{Zhuang}, \binits{Y.}},
\oauthor{\bsnm{Hua}, \binits{L.}},
\oauthor{\bsnm{Qi}, \binits{L.}},
\oauthor{\bsnm{Yang}, \binits{J.}},
\oauthor{\bsnm{Cao}, \binits{P.}},
\oauthor{\bsnm{Cao}, \binits{Y.}},
\oauthor{\bsnm{Wu}, \binits{Y.}},
\oauthor{\bsnm{Thompson}, \binits{J.}},
\oauthor{\bsnm{Haas}, \binits{H.}}:
A survey of positioning systems using visible led lights.
IEEE Communications Surveys and Tutorials
\textbf{20}
(2018).
\doiurl{10.1109/COMST.2018.2806558}
\end{botherref}
\endbibitem

\bibitem{Obeed2019}
\begin{botherref}
\oauthor{\bsnm{Obeed}, \binits{M.}},
\oauthor{\bsnm{Salhab}, \binits{A.M.}},
\oauthor{\bsnm{Alouini}, \binits{M.S.}},
\oauthor{\bsnm{Zummo}, \binits{S.A.}}:
On optimizing vlc networks for downlink multi-user transmission: A survey.
IEEE Communications Surveys and Tutorials
\textbf{21}
(2019).
\doiurl{10.1109/COMST.2019.2906225}
\end{botherref}
\endbibitem

\bibitem{Parker}
\begin{botherref}
\oauthor{\bsnm{Leung-Yan-Cheong}, \binits{S.K.}},
\oauthor{\bsnm{Hellman}, \binits{M.E.}}:
The gaussian wire-tap channel.
IEEE Transactions on Information Theory
\textbf{24}
(1978).
\doiurl{10.1109/TIT.1978.1055917}
\end{botherref}
\endbibitem

\bibitem{Oggier2011}
\begin{botherref}
\oauthor{\bsnm{Oggier}, \binits{F.}},
\oauthor{\bsnm{Hassibi}, \binits{B.}}:
The secrecy capacity of the mimo wiretap channel.
IEEE Transactions on Information Theory
\textbf{57}
(2011).
\doiurl{10.1109/TIT.2011.2158487}
\end{botherref}
\endbibitem

\bibitem{Classen2015TheSN}
\begin{botherref}
\oauthor{\bsnm{Classen}, \binits{J.}},
\oauthor{\bsnm{Chen}, \binits{J.}},
\oauthor{\bsnm{Steinmetzer}, \binits{D.}},
\oauthor{\bsnm{Hollick}, \binits{M.}},
\oauthor{\bsnm{Knightly}, \binits{E.}}:
The spy next door: Eavesdropping on high throughput visible light
  communications.
Proceedings of the 2nd International Workshop on Visible Light Communications
  Systems
(2015)
\end{botherref}
\endbibitem

\bibitem{Wang2018Physical-layerAnalysis}
\begin{barticle}
\bauthor{\bsnm{Wang}, \binits{J.Y.}},
\bauthor{\bsnm{Liu}, \binits{C.}},
\bauthor{\bsnm{Wang}, \binits{J.B.}},
\bauthor{\bsnm{Wu}, \binits{Y.}},
\bauthor{\bsnm{Lin}, \binits{M.}},
\bauthor{\bsnm{Cheng}, \binits{J.}}:
\batitle{{Physical-layer security for indoor visible light communications:
  Secrecy capacity analysis}}.
\bjtitle{IEEE Transactions on Communications}
\bvolume{66}(\bissue{12}),
\bfpage{6423}--\blpage{6436}
(\byear{2018}).
\doiurl{10.1109/TCOMM.2018.2859943}
\end{barticle}
\endbibitem

\bibitem{Lapidoth1998TheChannel}
\begin{botherref}
\oauthor{\bsnm{Lapidoth}, \binits{A.}},
\oauthor{\bsnm{Shitz}, \binits{S.S.}}:
{The Poisson Multiple-Access Channel}
\textbf{44}(2),
488--501
(1998)
\end{botherref}
\endbibitem

\bibitem{Strogatz2018}
\begin{bbook}
\bauthor{\bsnm{Strogatz}, \binits{S.H.}}:
\bbtitle{{Nonlinear Dynamics and Chaos}}.
\bpublisher{CRC Press},
\blocation{Boca Raton}
(\byear{2015}).
\doiurl{10.1201/9780429492563}
\end{bbook}
\endbibitem

\bibitem{XiaogeWu2016}
\begin{bchapter}
\bauthor{\bsnm{Wu}, \binits{X.}},
\bauthor{\bsnm{Zhang}, \binits{L.}}:
\bctitle{A l-i-oscm mapping method for improving ber of chaotic cdma based vlc
  system}.
In: \bbtitle{2016 IEEE Advanced Information Management, Communicates,
  Electronic and Automation Control Conference (IMCEC)},
pp. \bfpage{1028}--\blpage{1032}
(\byear{2016}).
\doiurl{10.1109/IMCEC.2016.7867367}
\end{bchapter}
\endbibitem

\bibitem{Ryu2013}
\begin{bchapter}
\bauthor{\bsnm{Ryu}, \binits{H.G.}},
\bauthor{\bsnm{Lee}, \binits{J.H.}}:
\bctitle{{High security wireless CDSK-based chaos communication with new chaos
  map}}.
In: \bbtitle{Proceedings - IEEE Military Communications Conference MILCOM},
pp. \bfpage{786}--\blpage{790}
(\byear{2013}).
\doiurl{10.1109/MILCOM.2013.139}
\end{bchapter}
\endbibitem

\bibitem{Sadoudi2013}
\begin{bchapter}
\bauthor{\bsnm{Sadoudi}, \binits{S.}},
\bauthor{\bsnm{Tanougast}, \binits{C.}},
\bauthor{\bsnm{Azzaz}, \binits{M.S.}}:
\bctitle{A new robust additive hyperchaos masking algorithm for secure digital
  communications}.
In: \bbtitle{2013 International Conference on Control, Decision and Information
  Technologies (CoDIT)},
pp. \bfpage{501}--\blpage{504}
(\byear{2013}).
\doiurl{10.1109/CoDIT.2013.6689595}
\end{bchapter}
\endbibitem

\bibitem{Fu2015}
\begin{bchapter}
\bauthor{\bsnm{Fu}, \binits{C.}},
\bauthor{\bsnm{Lin}, \binits{Y.}},
\bauthor{\bsnm{Jiang}, \binits{H.-y.}},
\bauthor{\bsnm{Ma}, \binits{H.-f.}}:
\bctitle{Medical image protection using hyperchaos-based encryption}.
In: \bbtitle{2015 9th International Symposium on Medical Information and
  Communication Technology (ISMICT)},
pp. \bfpage{103}--\blpage{107}
(\byear{2015}).
\doiurl{10.1109/ISMICT.2015.7107507}
\end{bchapter}
\endbibitem

\bibitem{Hammami2016}
\begin{bchapter}
\bauthor{\bsnm{Hammami}, \binits{S.}},
\bauthor{\bsnm{Djemaï}, \binits{M.}},
\bauthor{\bsnm{Busawon}, \binits{K.}}:
\bctitle{Encrypted audio communication design using synchronized discrete-time
  hyperchaotic maps}.
In: \bbtitle{2016 10th International Symposium on Communication Systems,
  Networks and Digital Signal Processing (CSNDSP)},
pp. \bfpage{1}--\blpage{6}
(\byear{2016}).
\doiurl{10.1109/CSNDSP.2016.7574015}
\end{bchapter}
\endbibitem

\bibitem{Yang2015}
\begin{bchapter}
\bauthor{\bsnm{Yang}, \binits{N.}},
\bauthor{\bsnm{Wu}, \binits{C.}},
\bauthor{\bsnm{Liu}, \binits{C.}},
\bauthor{\bsnm{Liu}, \binits{K.}}:
\bctitle{Adaptive synchronization of a novel fractional-order hyperchaotic
  system with uncertain parameters}.
In: \bbtitle{2015 IEEE 7th International Conference on Cybernetics and
  Intelligent Systems (CIS) and IEEE Conference on Robotics, Automation and
  Mechatronics (RAM)},
pp. \bfpage{248}--\blpage{252}
(\byear{2015}).
\doiurl{10.1109/ICCIS.2015.7274629}
\end{bchapter}
\endbibitem

\bibitem{Gunawan2020}
\begin{botherref}
\oauthor{\bsnm{Gunawan}, \binits{W.H.}},
\oauthor{\bsnm{Liu}, \binits{Y.}},
\oauthor{\bsnm{Yeh}, \binits{C.H.}},
\oauthor{\bsnm{Chow}, \binits{C.W.}}:
{Color-shift-keying embedded direct-current
  optical-orthogonal-frequency-division-multiplexing (CSK-DCO-OFDM) for visible
  light communications (VLC)}.
IEEE Photonics Journal
\textbf{12}(5)
(2020).
\doiurl{10.1109/JPHOT.2020.3021032}
\end{botherref}
\endbibitem

\bibitem{Chen2016}
\begin{botherref}
\oauthor{\bsnm{Chen}, \binits{B.}},
\oauthor{\bsnm{Zhang}, \binits{L.}},
\oauthor{\bsnm{Lu}, \binits{H.}}:
{High Security Differential Chaos-Based Modulation With Channel Scrambling for
  WDM-Aided VLC System}.
IEEE Photonics Journal
\textbf{8}(5)
(2016).
\doiurl{10.1109/JPHOT.2016.2607689}
\end{botherref}
\endbibitem

\bibitem{Fang2017}
\begin{bchapter}
\bauthor{\bsnm{Fang}, \binits{F.}},
\bauthor{\bsnm{Zhang}, \binits{L.}},
\bauthor{\bsnm{Rao}, \binits{W.}},
\bauthor{\bsnm{Chen}, \binits{W.}}:
\bctitle{Real-valued gram-schmidt transform-based non-coherent chaos shift
  keying scheme for visible light communication system}.
In: \bbtitle{2017 IEEE/CIC International Conference on Communications in China
  (ICCC)},
pp. \bfpage{1}--\blpage{6}
(\byear{2017}).
\doiurl{10.1109/ICCChina.2017.8330401}
\end{bchapter}
\endbibitem

\bibitem{Richter2002TheChaos}
\begin{barticle}
\bauthor{\bsnm{Richter}, \binits{H.}}:
\batitle{{The generalized H{\'{e}}non maps: Examples for higher-dimensional
  chaos}}.
\bjtitle{International Journal of Bifurcation and Chaos in Applied Sciences and
  Engineering}
\bvolume{12}(\bissue{6}),
\bfpage{1371}--\blpage{1384}
(\byear{2002}).
\doiurl{10.1142/S0218127402005121}
\end{barticle}
\endbibitem

\bibitem{AbdulRahman2019SecureSystems}
\begin{botherref}
\oauthor{\bsnm{Abdul~Rahman}, \binits{S.}},
\oauthor{\bsnm{Zribi}, \binits{M.}},
\oauthor{\bsnm{Smaoui}, \binits{N.}}:
{Secure communications based on the projective synchronization of
  four-dimensional hyperchaotic systems}.
Mathematical Problems in Engineering
\textbf{2019}
(2019).
\doiurl{10.1155/2019/2491850}
\end{botherref}
\endbibitem

\bibitem{Rodriguez2013}
\begin{botherref}
\oauthor{\bsnm{Rodr{\'{i}}guez}, \binits{S.P.}},
\oauthor{\bsnm{Jim{\'{e}}nez}, \binits{R.P.}},
\oauthor{\bsnm{Mendoza}, \binits{B.R.}},
\oauthor{\bsnm{Hern{\'{a}}ndez}, \binits{F.J.L.}},
\oauthor{\bsnm{Alfonso}, \binits{A.J.A.}}:
{Simulation of impulse response for indoor visible light communications using
  3D CAD models}.
EURASIP Journal on Wireless Communications and Networking
\textbf{2013}(1)
(2013).
\doiurl{10.1186/1687-1499-2013-7}
\end{botherref}
\endbibitem

\bibitem{Komine2004}
\begin{botherref}
\oauthor{\bsnm{Komine}, \binits{T.}},
\oauthor{\bsnm{Nakagawa}, \binits{M.}}:
{Fundamental analysis for visible-light communication system using LED lights}.
IEEE Transactions on Consumer Electronics
\textbf{50}(1)
(2004).
\doiurl{10.1109/TCE.2004.1277847}
\end{botherref}
\endbibitem

\end{thebibliography}


\end{document}